\documentclass[aps,prl,twocolumn,showpacs,floatfix,superscriptaddress,groupedaddress,preprintnumbers]{revtex4}

\usepackage{alltt}  

\usepackage{graphicx}

\pdfoutput=1

\usepackage{color}
\definecolor{linkcolor}{RGB}{0,0,128}
\usepackage[pdftex,pdfpagemode={UseOutlines},bookmarks,bookmarksopen,colorlinks,linkcolor=linkcolor,citecolor={black},urlcolor=linkcolor]{hyperref}
\usepackage{bm}
\usepackage{amsmath,amssymb}
\usepackage{aas_macros}
\usepackage{url}
\usepackage[mathscr]{eucal}

\newcommand{\photoz}{photo-$z$}

\newcommand{\wx}{W_{X}}

\newcommand{\wy}{W_{Y}}

\newcommand{\thetab}{\boldsymbol\theta}
\newcommand{\weight}{\mathtt{w}}

\newcommand{\psiclust}{\psi^{\rm clust}}

\newcommand{\psilens}{\psi^{{\rm lens}}}

\newcommand{\optbasishat}{\hat{\varphi}^{\rm opt}}

\begin{document}

\hspace{5.2in} \mbox{LLNL-JRNL-644558}

\title{Probing Dark Energy with Lensing Magnification in Photometric Surveys}
\author{Michael D. Schneider}

\affiliation{Lawrence Livermore National Laboratory, P.O. Box 808 L-210, Livermore, CA 94551-0808, USA.}
\affiliation{University of California, One Shields Avenue, Davis, CA 94551, USA.}

\begin{abstract}
I present an estimator for the angular cross-correlation of two 
tracers of the cosmological large-scale structure that utilizes 
redshift information to isolate separate physical contributions.
The estimator is derived by solving the Limber equation 
for a re-weighting of the foreground tracer that 
nulls either clustering or lensing 
contributions to the cross-correlation function. 
Applied to future photometric surveys, the estimator 
can enhance the measurement of gravitational 
lensing magnification effects to provide a
competitive independent constraint on the dark energy equation of state. 
\end{abstract}


\maketitle

\paragraph{Introduction}
The two-point correlation functions of galaxies and quasars have proven 
to be powerful probes of cosmological models~\cite[e.g.][]{wigglez}. 
Over the next two 
decades, wide-field astronomical 
surveys~\footnote{\url{http://www.darkenergysurvey.org},
\url{http://www.naoj.org/Projects/HSC/},
\url{http://kids.strw.leidenuniv.nl/},
\url{http://www.lsst.org/},
\url{http://sci.esa.int/euclid/}, 
\url{http://wfirst.gsfc.nasa.gov/}} 
 will rely on two-point correlation functions to self-calibrate 
several types of
systematic error~\cite[e.g.][]{schneider06, zhang08, bernstein09, mandelbaum12} 
and constrain models of dark energy~\cite[e.g.][]{mandelbaum12, descwp, basse13}.

Precise interpretation of correlation function measurements requires 
knowledge of the line-of-sight distribution of the sources. Astronomical 
surveys obtain this information either with a spectrograph, which requires
long integration times and therefore limits the total number of sources 
observed, or with broad-band photometry, which allows improved statistical 
precision at the expense of larger errors in the line-of-sight 
source distribution.

In this letter, I derive an estimator for cosmological cross-correlation
functions that simultaneously removes the ambiguities in theoretical 
interpretation when the line-of-sight source distributions have large 
observational errors and maximizes the signal-to-noise ratio (SNR) 
of shot-noise limited measurements through optimal sample selection.
The method of Ref.~\cite{seljak09} is complimentary to this letter 
in optimizing the SNR for sample variance dominated measurements. 

While the cross-correlation estimator is broadly applicable, 
I will focus on the measurement of 
gravitational lensing magnification, where an optimal 
estimator promises particularly large scientific gains.
Gravitational lensing by cosmological large-scale structure causes 
the images of background sources to be both magnified and sheared. 
Lensing magnification alters the apparent number density of background 
sources by two competing effects, 
1) magnifying the area in a given patch of sky thereby reducing the 
source number density, and 
2) increasing the apparant brightness of sources otherwise just below a survey detection 
limit therby increasing the observed source number density.
The dominant effect is determined by the slope of the differential source number counts 
as a function of magnitude, with steep slopes yielding increases in the apparant number 
density.

Most detections of lensing magnification to-date have measured the angular 
cross-correlation function of two source samples widely separated in 
redshift~\cite{scranton05, wang11, morrison12} 
(although see a novel method in Ref.~\cite{huff11}). 
In the absence of lensing, such a correlation would be zero, 
yielding a clean measurement of lensing effects. 

The cross-correlation of galaxy shapes to detect lensing shears (cosmic shear) 
is well established as an important cosmological probe~\cite[e.g.][]{jee13}.
Current and next generation surveys targeted at measuring dark energy properties via 
cosmic shear will rely on 
`photometric redshifts' (\photoz's) that measure a galaxy spectrum in a 
few broad-band optical filters. While \photoz's show 
great promise for statistical measurements of large-scale structure, the large 
\photoz~errors pose several challenges for analysis. The detection of lensing 
magnification through cross-correlation of \photoz~bins is contaminated by 
the intrinsic clustering of galaxies that are co-located in redshift 
and poorly separated due to the \photoz~errors. 
This difficulty, along with the intrinsically lower-amplitude correlations, have
made lensing magnification a less 
attractive probe of dark energy properties than
cosmic shear in photometric surveys,
although magnification has been appreciated 
as a means of self-calibrating systematic 
errors~\cite{zentner08,bernstein09,vallinotto11,eifler13}.

Consider then the two-point cross-correlation function of two tracers of the cosmological mass density
that are separated, but may partially overlap, in redshift.
The observed cross-correlation is the sum of 
those from intrinsic clustering where the samples overlap in 
redshift ($w_{gg}$), 
lensing magnification ($\mu$) of background sources by the mass around 
foreground galaxies ($w_{g\mu}$), and 
the lensing two-point correlation from line-of-sight structures 
that magnify both 
tracer samples ($w_{\mu\mu}$)~\citep{moessner98}.

The angular cross-correlation function of two catalogs can be estimated 
by summing the number of source pairings in angular bins ($\theta$) 
normalized to the 
expected pairings for catalogs with uniform positions over the 
sky~\cite{landy93},
\begin{equation}\label{eq:correstimator}
  \hat{w}(\theta) = \sum_{i=1}^{N_1} \sum_{j=1}^{N_2} 
  \weight_{i} \weight_{j} 
  \Delta_{\theta}\left(\left|\thetab_i-\thetab_j\right|\right) 
  / N_{\rm rand-pairs}(\theta; \weight)
\end{equation}
where $\Delta_{\theta}$ is one if the magnitude of the angular separation vector 
of source $i$ and source $j$ is in the bin centered at $\theta$ and 
$\weight_{i,j}$ are arbitrary weights for each pair of sources.

We seek a set of weights, $\weight_{i,j}$,
in the sum over pairs in the cross-correlation function estimator that will minimize 
or maximize the amplitude of either the clustering ($w_{gg}$) 
or lensing contributons ($w_{g\mu}$)
to the observed cross-correlation. 
Ref.~\cite{heavens11} pursued a related goal in attempting to minimize 
both $w_{gg}$ and $w_{g\mu}$ to measure $w_{\mu\mu}$. This latter term 
is typically an order-of-magnitude smaller than even $w_{g\mu}$, which would 
require unrealistically large galaxy samples for a statistically significant 
measurement. 
Ref.~\cite{zhang05} showed that the clustering ($w_{gg}$) and magnification
($w_{g\mu}$) terms can be separated using the strong luminosity dependence 
of the lensing magnification, but assuming sources can be cleanly separated 
into non-overlapping redshift bins. 
I take a more general approach for arbitrary cross-correlations that may 
include large uncertainties in the redshift distributions. 
I expect luminosity information will only further improve the method 
in this letter for the measurement of lensing magnification.

\paragraph{Derivation of the optimal estimator}

Each term contributing to the angular cross-correlation of cosmological 
tracers can be written as a projection of the 3D 
matter or galaxy correlation function, assuming 
Limber's approximation~\cite{limber53}, 
flat-sky approximation, and zero spatial curvature,
\begin{equation}\label{eq:angcorr}
  w_{XY}(\theta) = \int_{0}^{\chi_{\infty}} d\chi\, W_{X}(\chi) K(\chi,\theta),
\end{equation}
where $W_{X}(\chi)$ is either a redshift distribution
or lensing kernel~\cite[see the appendix of][]{moessner98} for a given sample,  
$K(\chi,\theta)\equiv W_{Y}(\chi)\xi(\chi\theta)$ is the product of the weight function 
for the background sample and the 3D galaxy correlation function,
\begin{equation}\label{eq:corr3d}
  \xi(r) \equiv \int \frac{k\,dk}{2\pi} P(k)\, J_{0}(kr),
\end{equation} 
$P(k)$ is the 3D matter or galaxy power spectrum, $J_{0}$ is the zeroth order 
Bessel function and we have neglected redshift space distortions. 

Eqn.~(\ref{eq:angcorr}) is a Fredholm integral equation of the first kind
 and can be solved for
$\wx(\chi)$ to yield a minimum or maximum angular correlation $w_{XY}(\theta)$.
At least for the standard cosmological model the source kernel $K(\chi,\theta)$
has a non-trivial null space such that there are non-trivial $W_{X}(\chi)$ 
that satisfy $\int d\chi\, W_{X}(\chi)K(\chi,\theta)=0$.
We will use functions in the null space of $K(\chi,\theta)$, 
assuming $W_{Y}(\chi)$ is either a lensing or clustering window function,
to construct 
foreground weights $W_{X}(\chi)$ that minimize the amplitude 
of $w_{XY}(\theta)$ for all $\theta$. 

The eigenvectors of the source kernel provide a convenient 
means to define an orthogonal basis in the range of the kernel operator.
The kernel $K(\chi,\theta)$ as defined by Eq.~\ref{eq:angcorr} is 
not Hermitian ($K(\chi,\theta) \neq K(\theta,\chi)$) due to the presence of 
$\wy(\chi)$ and we cannot be guaranteed to have real eigenvalues and 
orthogonal eigenvectors.
A better kernel is found by considering the square of $K(\chi,\theta)$,
\begin{equation}\label{eq:symkernel}
  C(\chi,\chi') \equiv \int d\theta\, K(\chi,\theta)\, K(\chi', \theta),
\end{equation}
with eigenvalue equation,
\begin{equation}\label{eq:KHeigen}
  \int d\chi\, C(\chi,\chi')\, \psi(\chi) = \lambda \psi(\chi').
\end{equation}
The kernel $C(\chi,\chi')$ is symmetric and normalizable, 
yielding orthogonal and non-trivial eigenvectors that can be used 
to construct a desireable solution for $W_{X}(\chi)$.

With the aid of the auxiliary symmetric kernel eigenvectors,
\begin{align}
  C_{\rm aux}(\theta,\theta') &\equiv \int d\chi\,
  K(\chi,\theta)\, K(\chi, \theta'), \notag\\
  \int d\chi\, C_{\rm aux}(\theta,\theta')\, \eta(\theta) &= \lambda \eta(\theta'),
\end{align}
the original source kernel can be reconstructed~\cite{kornkorn},
\begin{equation}
  K(\chi,\theta) = \sum_{i=1}^{\infty} \sqrt{\lambda_{i}}
  \hat{\psi}_{i}(\chi)\hat{\eta}_{i}(\theta),
\end{equation}
where hats denote unit normalized eigenvectors with an L2 norm.
So, solutions of the symmetric kernel eigenvalue equation also provide 
solutions for the original source kernel eigenvalue equation.

The algorithm for optimizing the cross-correlation estimator has 
three steps:
\begin{enumerate}
  \item Solve for the eigenfunctions of the lensing and clustering symmtric source kernels and define which physical effect is to be maximized (e.g. lensing) and which is to be nulled (e.g. clustering).
  \item Find components of the eigenfunctions of the
  source kernel to be maximized (e.g. lensing) that are in the 
  null space of the  source kernel to be nulled (e.g. clustering) 
  using Gram-Schmidt orthogonalization. Construct a basis set from these 
  components for the final weight functions.
  \item Solve for a combination of basis functions that optimizes the 
  signal-to-noise ratio to construct
  the pair weights for the cross-correlation function estimator in 
  Eqn.~\ref{eq:correstimator}. 
\end{enumerate}
The dominant Poisson contribution to the covariance of the angular correlation function 
is nearly diagonal, yielding a simple expression for the signal-to-noise 
ratio to optimize when we neglect sample variance,
\begin{equation}
  {\rm SNR} = \frac{1}{N_{X, \rm{eff}}(\weight) N_Y}
  \sum_{i} w(\theta_i; \weight)^2 \Delta\theta_i,
\end{equation}
where $\Delta\theta_i$ is the width of angular bin $i$, $N_Y$ is the 
number of sources in catalog $Y$, and the effective number of foreground 
sources in the catalog $X$ after re-weighting is,
\begin{equation}
  N_{X, \rm{eff}}(\weight) = \int d\chi\, \weight^{2}(\chi) \frac{dN_X}{d\chi},
\end{equation}
with the convention $-1 \le \weight \le 1$.

\begin{figure}
  \centerline{
    \includegraphics[width=0.45\textwidth]{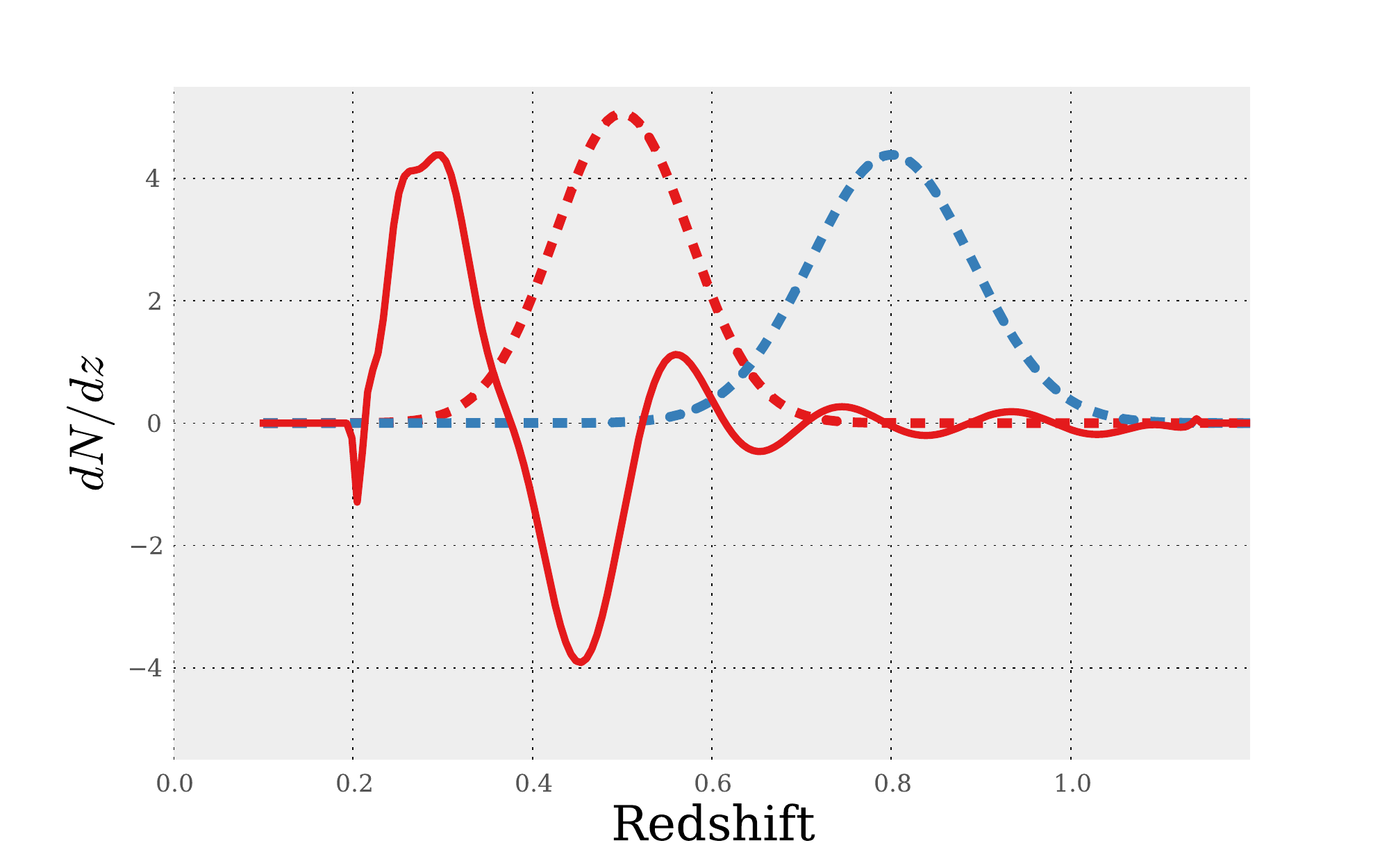}
  }
  \caption{Model foreground (red, dashed) and background (blue, dashed) 
  redshift distributions
  and the optimal reweighting of the foreground distribution (solid) to 
  minimize intrinsic clustering and optimize lensing in the cross-correlation 
  of the two redshift bins.
  }
  \label{fig:basisfunctions}
\end{figure}

\paragraph{Implementation and example}
I will describe the implementation of the optimal cross-correlation estimator 
by means of an example lensing magnification measurement that could be made with 
the Large Synoptic Survey Telescope (LSST)~\footnote{\url{http://lsst.org}}.
The LSST will measure $\sim10^9$ galaxies with \photoz\, rms error 
of $0.05(1 + z)$~\cite{lsst_sb}.
I consider 4 tomographic bins with centers 
linearly spaced between $z=0.5$ and $z=1.4$ (where the time-dependence of dark energy 
may be most easily detected). The redshift distributions of the 
first 2 tomographic bins are depicted by the dashed 
lines in Fig.~\ref{fig:basisfunctions}.
This bin spacing is a tradeoff between clean separation in redshift and 
informative sampling in redshift.
I consider only linear clustering predictions for the dark matter, with a maximum 
wavenumber in the matter power spectrum of $0.1\,h$Mpc$^{-1}$ and a 
multiplicative linear galaxy clustering bias relating the galaxy and matter 
power spectra, 
$P_{gg}(k; z_1,z_2) \equiv b_g(z_1)b_g(z_2)P_{mm}(k; z_1,z_2)$.

The optimal pair weights need to be recomputed for every 
pair of tomographic redshift bins. For two distinct bins 
in redshift with source distributions $n_1(z)$ and $n_2(z)$,
we first solve for the eigenfunctions
$\psilens_i(\chi)$ of the symmetric kernel
$
  C_{\rm lens}(\chi,\chi') \equiv 
  W_{Y,{\rm lens}}(\chi)W_{Y,{\rm lens}}(\chi')
  \int d\theta\, \xi(\chi\theta)\xi(\chi'\theta)
$
as well as the eigenfunctions $\psiclust_i(\chi)$ for the 
analogous clustering kernel.

Then, to down-weight the intrinsic clustering contribution to the 
final cross-correlation we select the components of 
$\psilens_i(\chi)$ that are orthogonal (with an L2 norm) 
to all the $\psiclust_{i}(\chi)$ for 
all $i$ up to a pre-specified numerical tolerance in the rank-ordered 
eigenvalue spectrum. 
After projection onto the null space of the clustering kernel the 
eigenfunctions are no longer orthogonal. Let 
$\optbasishat_{i}(\chi)$ denote the orthonormal basis constructed from the 
projected eigenfunctions via Gram-Schmidt orthogonalization.

We then define the optimal weights as functions of the line-of-sight distance, 
$\chi$, as,
\begin{equation}\label{eq:wxoptbasis}
  \wx(\chi) = \sum_{i=1}^{n_{\rm basis-funcs}} w_{i}\optbasishat_{i}(\chi)
  + \phi(\chi),
\end{equation}
where $\phi(\chi)$ is an arbitrary function in the null space of both 
the lensing and clustering kernels that can be used to minimize the shot noise 
and $w_i$ are parameters that we will set to optimize the signal-to-noise 
ratio of the correlation function measurement.

Most solutions for the optimal window will increase the shot noise 
by down-weighting redshift ranges where the galaxy number density is 
large. 
Intuitively, the shot noise should be minimized when the optimal 
window is similar to the observed redshift distribution of the 
foreground sample (so that all galaxies have nearly equal magnitude 
weights). A useful guess for $\phi(\chi)$ in Eq.~\ref{eq:wxoptbasis}
could then be the original foreground redshift distribution projected 
into the null space of the lensing and clustering kernels.

I assume in Eq.~\ref{eq:wxoptbasis} that the basis functions are sorted in order 
of decreasing eigenvalue of the kernel to be maximized (e.g. lensing). 
The total number of basis functions in Eq.~\ref{eq:wxoptbasis} is limited 
in practice by numerical errors in the discretization of the continuous 
eigenvalue Eq.~\ref{eq:KHeigen}. I use 20 Gauss quadrature weights for the 
example pair of bins in Fig.~\ref{fig:basisfunctions}. 
By calculating basis vectors of the 
kernel null spaces and then applying the kernel to these basis functions (which 
should yield zeros for all $\chi$) I find the first 4 eigenfunctions yield 
numerical errors at least 3 orders of magnitude below the expected amplitudes of the 
lensing cross-correlations.
The solid red line in Fig.~\ref{fig:basisfunctions} shows the solution 
from Eq.~\ref{eq:wxoptbasis} for the 2-bin example. 
For this optimal weighting, $9\times10^5$ galaxies are needed in the foreground 
redshift bin to obtain a signal-to-noise ratio of one.

\begin{figure}
  \centerline{
    \includegraphics[width=0.45\textwidth]{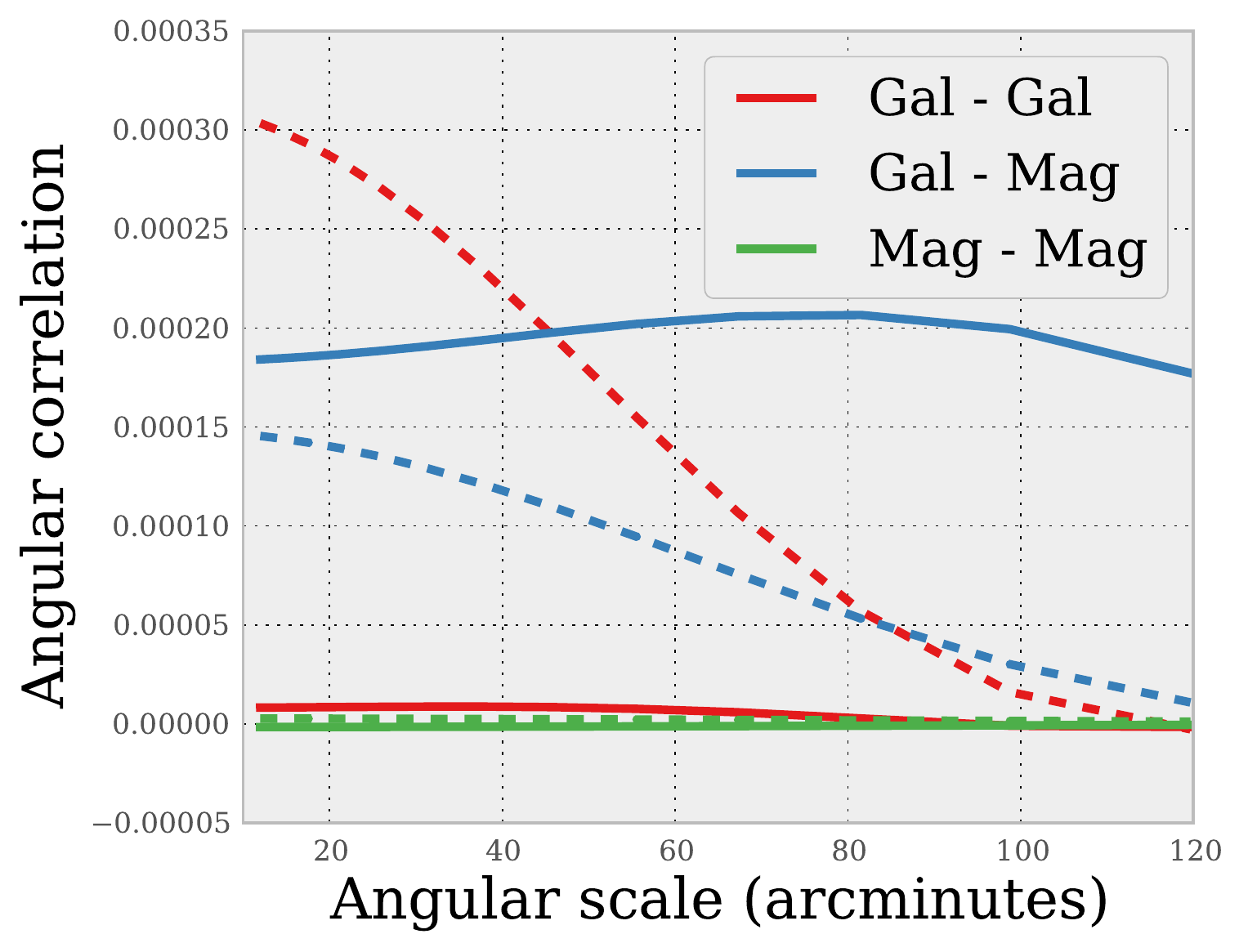}
  }
  \caption{Cross-correlation functions between two photometric redshift bins 
  with (solid) and without (dashed) optimal weighting. 
  The contributions to the cross-correlation include galaxy clustering ('Gal') 
  and lensing magnification ('Mag').
  The optimal weights both reduce the 
  intrinsic clustering amplitude and increase the amplitude of the lensing 
  magnification cross-correlation.}
  \label{fig:crosscorr}
\end{figure}

In Fig.~\ref{fig:crosscorr} I show the predicted components of the angular
cross-correlation function for the 2 redshift bins shown in 
Fig.~\ref{fig:basisfunctions} assuming the uniform foreground weighting 
(dashed) and the optimal weighting (solid). Using the optimal window 
reduces the amplitude of the intrinsic galaxy clustering correlation (red)
(due to the partial overlap of the redshift bins) by an order of magnitude and
increases the amplitude of the magnification-clustering cross-correlation (blue)
over all scales considered. Nulling the lensing magnification instead reduces 
all the lensing contributions to at least 2 orders of magnitude below the 
intrinsic clustering.

\begin{figure}
  \centerline{
    \includegraphics[width=0.45\textwidth]{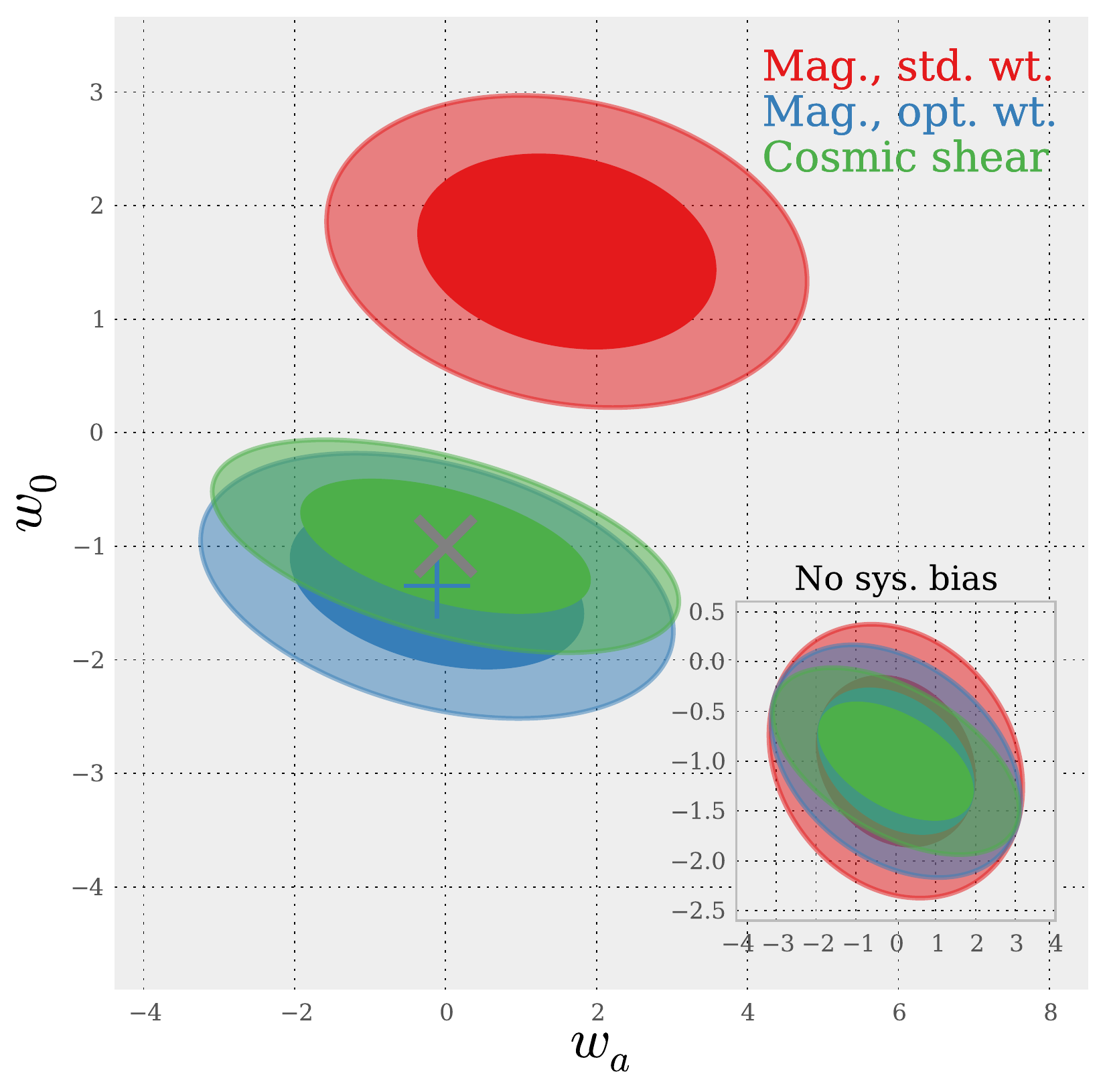}
  }
  \caption{Forecasted constraints on dark energy equation of state parameters 
  from galaxy angular auto- and cross-correlation functions in 4 tomographic bins 
  limited to the 
  linear clustering regime. 
  Inset panel: no systematic biases. Main panel: assuming an uncorrected 
  1\% systematic bias in the linear galaxy clustering bias.
  The largest (red) contour shows the constraints using 
  uniform weighting for each galaxy pair in the correlation function 
  estimator. The smaller contours (blue) show the constraints when using the 
  optimal redshift weighting in this letter. For comparison, the smallest contours 
  (green) show the constraints with a cosmic shear measurement (correlated galaxy 
  shapes). 
  In each case, I marginalize over uncertainties in the \photoz\, 
  distributions in each bin (mean, variance, and outlier fraction) as well 
  as the linear clustering bias in each bin, with 30\% priors. 
  The optimal weighting in the magnification cross-correlations increases the 
  dark energy Figure of Merit
  by 25\% over uniform galaxy weighting and reduces 
  the systematic parameter biases to less than the 1-$\sigma$ uncertainties.
  }
  \label{fig:contours}
\end{figure}

In Fig.~\ref{fig:contours} I show forecasted constraints on the dark energy 
equation of state parameters $w_0$ and $w_a$, where the 
time-dependent equation of state is modeled as, 
$w(a) = w_0 + (1-a)w_a$~\cite{linder05}.
Using the Fisher matrix, 
I forecast 3 scenarios that might be measured 
with LSST: 
1) Galaxy cross-correlations using the 4 tomographic bins described above and 
uniform weights in the correlation estimator,
2) Galaxy cross-correlations with optimal weighting to amplify, in turn, the 
lensing and clustering contributions,
3) Cosmic shear measured with the (cross-)correlation of galaxy shapes.
In all scenarios I marginalize over cosmological parameters 
$\sigma_8$ and $\Omega_m$, 3 \photoz\, error 
parameters (the \photoz\, bin mean, bin variance, and `catastrophic' outlier 
fraction), and the linear clustering bias for scenarios 1 and 2. I assume 
Planck priors~\cite{planck13} on the cosmological parameters and nuisance parameter priors 
derived from the LSST tolerances~\cite{lsst_sb}. 
The Planck prior dominates 
the constraints on the $w_a$ axis and alleviates sensitivities to the 
choice of fiducial \photoz\, error model.

The contour offsets in Fig.~\ref{fig:contours} show the parameter biases
when the linear galaxy clustering bias is systematically mis-estimated by 1\% (the inset 
assumes zero systematic bias). 
Such systematics are major challenges for the conventional galaxy clustering measurement, 
but are self-calibrated with the optimally weighted clustering measurement. 

\paragraph{Conclusions}
Figure~\ref{fig:contours} demonstrates that, with optimization, the 
lensing magnification measured via cross-correlations 
of \photoz~bins may yield a dark energy Figure of Merit 
(defined as the inverse of the $w_0-w_a$ ellipse area)~\cite{detf} up to 
80\% that from cosmic shear with the same set of galaxies and 
tomographic binning. Because the lensing magnification only depends 
on the clustering bias of the foreground sample, the optimally weighted 
cross-correlations are able to partially self-calibrate an 
uncorrected systematic error in the bias, further improving the 
dark energy measurement relative to the standard redshift weighting.
By down-weighting the lensing contributions to 
cosmological angular cross-correlations with the same source catalog, 
we can also improve the measurement of 
the linear clustering bias with respect to the dark matter and the 
calibration of photometric redshift uncertainties~\cite{matthews10}.

For partially overlapping redshift bins such as those modeled in 
Fig.~\ref{fig:basisfunctions}, the optimal weighting effectively discards 
a large number of galaxies. We 
therefore require large surveys such as the LSST to reduce the shot noise. 
Typically, at least $10^6$ sources are required in each foreground redshift bin 
to detect lensing magnification via cross-correlations with our optimal 
redshift weighting. But LSST will yield at least 10 times more galaxies per bin 
near the peak of the redshift distribution, with the exact useable number depending 
on the accuracy of the photometric calibration. 

Other possible sources of systematic errors in the optimal redshift 
weighting include unknown variations of the source redshift distributions 
or luminosity functions due to photometric calibration errors 
over a survey footprint on the sky and uncertainties in the probability 
distribution for the true redshift of each foreground object. 
Because the optimal redshift weights are sensitive to the model \photoz\, 
distributions, an iterative solution for the redshift weights may yield 
important information on the \photoz\, calibration as well.

\acknowledgments
I thank Shaun Cole and Tony Tyson for helpful discussions.
This work performed under the auspices of the U.S. Department of Energy by Lawrence Livermore National Laboratory under Contract DE-AC52-07NA27344.

\bibliography{optimalEstimator}
\end{document}